\documentclass[authoryear,12pt,3p,1in]{jowarticle}

\usepackage[english]{babel}
\usepackage[utf8]{inputenc}
\usepackage{amsmath}
\usepackage{graphicx}
\usepackage{setspace}
\usepackage[colorinlistoftodos]{todonotes}
\usepackage{natbib}

\makeatletter
\renewcommand\section{\@startsection{section}{1}{\z@}{-3.25ex plus -1ex minus -.2ex}{1.5ex plus .2ex}{\normalsize\bf}}
\renewcommand\subsection{\@startsection{subsection}{2}{\z@}{-3.25ex plus -1ex minus -.2ex}{1.5ex plus .2ex}{\normalsize\bf}}
\renewcommand\subsubsection{\@startsection{subsubsection}{3}{\z@}{-3.25ex plus -1ex minus -.2ex}{1.5ex plus .2ex}{\normalsize\bf}}

\usepackage{enumerate}
\usepackage{graphicx}
\usepackage[round]{natbib}
\usepackage{amssymb,amsmath, amscd, amsthm, mathrsfs}

\usepackage{ bbold }
\usepackage{enumitem}
\setlist{nosep}
\usepackage{newtxtext}
\theoremstyle{definition}

\theoremstyle{remark}

\numberwithin{equation}{section}

\renewcommand{\footnotesize}{\normalsize}

\makeatother

\begin{document}
\begin{frontmatter}
\title{We Have Never Been Sophisticated}

\author{Clara Bradley}\ead{clara.bradley@ucl.ac.uk}
\address{Department of Philosophy \\ University College London}
\author{James Owen Weatherall}\ead{weatherj@uci.edu}
\address{Department of Logic and Philosophy of Science \\ University of California, Irvine}

\date{\today}

\begin{abstract}

Many philosophers of physics maintain that a physical theory that exhibits (certain kinds of) symmetries is flawed, on the grounds that such theories posit ``excess structure". In an influential paper, Dewar [2019, ``Sophistication about Symmetries'', \emph{Brit. J. Phil. Sci.} \textbf{70}: 485-521] introduces a distinction between ``reduction" and ``sophistication" as alternative ways of removing excess structure.  In this paper we re-examine the distinction as Dewar draws it, and we argue that there is no physically or philosophically important distinction between what Dewar calls ``reduction'' and what he calls ``internal sophistication''. We then argue that there are multiple notions of ``reduction" in the literature that ought to be distinguished, both in motivation and in outcome. 

\end{abstract}
\end{frontmatter}
\newpage
\onehalfspacing
\section{Introduction}\label{sec:intro}

In an influential recent paper, Neil \citet{DewarSoph} advances a doctrine that he dubs \emph{sophistication} about symmetries.\footnote{\label{fn:otherPeople} Since circulating an early version of this paper, several colleagues have suggested in conversation and correspondence that although Dewar introduced the terminology of ``internal sophistication'' and ``external sophistication'', his treatment is not the authoritative one, and that some other usage has been adopted in the subsequent literature, especially very recently \citep[e.g.][]{Gomes,ReadVIBES,MarchIntrinsic,Jacobs+March}.  Our focus here is on the (in our estimation, influential) distinction Dewar draws in the body of his paper.  In our view, more recent discussions either suffer from the same problems as Dewar's, or else simply change the subject.  We will substantiate this claim in footnote \ref{fn:relativity}, after we have laid out Dewar's views and our response, by discussing a recent attempt by \citet{March+Read} to give precise definitions of reduction and internal sophistication intended to track the more recent literature.}  Roughly, the reasoning goes as follows.  Many philosophers of physics maintain that when a physical theory exhibits symmetries, it is, \emph{ipso facto}, flawed.  This is because it is believed that a theory with symmetries posits or has or otherwise invokes ``excess'' structure, viz., whatever varies under application of the symmetry transformations.  (An effort to make these ideas more precise will be made below.)  Given such a theory, a common strategy is to seek out an alternative theory, empirically equivalent to the first, that does not have symmetries. This alternative theory, on Dewar's terminology, is said to be a \emph{reduced} theory, and the strategy just described is called \emph{reduction}.  But according to Dewar, reduction is neither the only nor the obviously best option.  Instead, he argues, one could also pursue \emph{sophistication}.  Rather than change the theory itself---that is, its syntax or formalism---one changes how one interprets the theory.\footnote{In the first instance, Dewar means ``formally interpret'', as in the sense of formal semantics. In what follows, we sometimes elide the difference between ``theory'' as a syntactic object and ``theory'' as that plus a (formal) interpretation.  For instance, we will use ``sophisticated theory'' as shorthand for ``theory formally interpreted in a sophisticated manner'', in contrast with ``reduced theories''. There is also a further issue regarding how to interpret theories as bearing on their worldly subject matter, which we return to in section \ref{sec:auxiliary}.  The differences in formal interpretation strategies that Dewar discusses are meant to bear on interpretation is this other sense, but how they do so is not our focus here.  (See \citet[\S 3]{Halvorson+etal} for a discussion of the relationship between formal interpretation and interpretation in the other sense.)}  On this strategy, one still denies physical significance to the symmetry-variant features of the theory, but one does not eliminate them from the theory's syntax.

Several authors have found advantages in sophistication, as compared to reduction.  Dewar himself, for instance, argues that sophisticated theories have explanatory advantages over their reduced alternatives, because of the additional resources afforded by the ``extra'' syntactic structure.\footnote{ See also \citet{BradleyLorentz}, who offers a more robust case in which theories with different (amounts of) structure provide very different explanations, despite being empirically equivalent.  For her, this means that one cannot always remove structure from a theory, even if it plays no role in the theory's predictions, without incurring other costs.}  \citet{BradleySoph}, meanwhile, has argued that sophisticated theories may be preferred because they have richer ``auxiliary'' structure: that is, they more readily allow one to represent reference systems, or observers, or relationships between subsystems.  In this way, sophisticated theories may have pragmatic advantages over reduced ones.  And perhaps most compelling of all, it is not clear that reduced theories exist in all cases---and even if they do, there is no known recipe to construct them.  Thus, at least sometimes, sophistication has all of the advantages of being the only available option.

These advantages notwithstanding, our goal in the present paper is to show that the reduction/sophistication distinction has not been cleanly drawn.  Dewar introduces sophistication first in the context of first-order model theory.  He offers a precise definition, in that context, of a ``symmetry of a theory'', and then he gives clear statements of what it means to move from a theory with symmetries to either a ``reduced'' or a ``sophisticated'' alternative.  He then distinguishes two approaches to sophistication, which he calls ``internal'' and ``external''.  (We discuss this distinction at length below.)  Dewar's discussion here is clear, and we take it to fix what the relevant terms are supposed to mean on his account.  Dewar then suggests that similar ideas, \emph{mutatis mutandis}, can be applied in the context of physical theories. (He focuses on what he calls ``local field theories'', including general relativity and classical Yang-Mills theory.)  Here, too, he introduces a notion of symmetry and suggests that reduction, internal sophistication, and external sophistication are all distinct and viable options, arguing largely by analogy with the first-order case.  

Dewar's treatment of sophistication takes inspiration---and terminology---from the literature on substantivalism and relationism in the context of general relativity.\footnote{ To be sure, Dewar acknowledges that general relativity is not the perfect example of what he has in mind, for reasons related to what we argue below.  But he also indicates that fiber bundle formulations of Yang-Mills theory, along the lines of those discussed by \citet{WeatherallFBYMGR}, would count as good examples of (internally) sophisticated theories, and these are saliently analogous to general relativity.} There, a family of related ``realist'' interpretations of the theory are often lumped together under the name ``sophisticated substantivalism''.  Sophisticated substantivalists are united by two claims: first, that spacetime is a physically real entity; and second, that distinct-but-isometric models of general relativity do not correspond to distinct possibilities.\footnote{ The expression ``sophisticated substantivalism'' was coined as a slur by \citet{Belot+Earman}, but quickly embraced by advocates for the view(s). There are different species of sophisticated substantivalism.  To sample from the more widely discussed variants: some are motivated by Lewisian counterpart theory \citep{Butterfield,Brighouse}; others by anti-haecceitism about spacetime points \citep{Hoefer,PooleyOld}; and yet others adopt further claims about the essential properties of spacetime points \citep{Maudlin}. Most species of sophisticated substantivalism are motivated by the hole argument \citep{Earman+Norton}, though recently \citet{WeatherallHoleArg} has argued that the hole argument is mathematically flawed, but nonetheless defends a view similar in many particulars to anti-haecceitist flavors of sophisticated substantivalism on other grounds.} It is the second of these claims that connects the view to Dewar's sophistication, as described above.  This is because the distinct-but-isometric models at issue there are often taken to be related by a ``diffeomorphism symmetry''.\footnote{ Whether this is even true is contentious---\citet{Weatherall+March} argue that on a natural definition of the ``symmetries of a theory'' for covariant field theories, diffeomorphisms are \emph{not} symmetries of general relativity.  To be sure, diffeomorphisms can be seen to generate isomorphisms between models of the theory, but they do not ``act on the theory'' in any coherent sense.}  Thus we have models that are distinct, at least at some level of description, and yet are related by a symmetry; and according to sophisticated substantivalism, we are meant to interpret general relativity in such a way that whatever distinguishes the models has no physical significance.  Indeed, \citet{DewarSoph} suggests that he intends (internal) ``sophistication'' in much the same way as it is used in philosophy of general relativity.  To be sophisticated about symmetries is to treat \emph{other} cases analogously to how sophisticated substantivalism treats diffeomorphisms in general relativity.

But as we will argue, there are formal mismatches between symmetries as Dewar defines them in the first-order case and those in the local field theory case, particularly with regard to the ``symmetries'' of interest to the sophisticated substantivalist.  The key issue is whether symmetries of a theory can act merely by permuting elements of a model's domain or, what amounts to the same thing, if symmetries can relate isomorphic models.\footnote{ There is a close connection, here, to arguments originally made by \citet{WeatherallUG}, which \citet{DewarSoph} discusses and almost endorses. On Weatherall's view, theories such that two models have the same representational capacities (if and) only if they are isomorphic should not be seen to exhibit excess structure, and thus are not ``gauge theories''; he also argues that one way to ``remove'' excess structure from a theory is to introduce, by formal stipulation, additional isomorphisms---what Dewar called external sophistication.}  In the first-order case, non-trivial symmetries cannot do either of these.  But in the field theoretic case, they can.  The result is that the distinction between sophisticated and reduced theories in the first-order case does not align well with the field theoretic case.  In fact, there are \emph{no} non-trivial reduced local field theories, if by ``reduced'' we mean a theory with no symmetries, and by ``symmetry'' we mean symmetries as Dewar defines them for field theories.  If, by contrast, one thinks of the analogy between first-order and local field theories somewhat differently, so that the resulting notion of symmetry shares the formal properties of symmetries as Dewar defines them in the first-order case, then not only do reduced theories abound, many of our standard formulations of theories---ones Dewar would call internally sophisticated, such as general relativity and fiber bundle formulations of Yang-Mills theory---should count as reduced.  We will conclude that there is no interesting distinction to be drawn between reduced theories and internally sophisticated ones for local field theories. At best, internally sophisticated theories should be seen as reduced theories with certain additional formal properties of unclear philosophical or physical significance.

This argument will raise questions for some readers, because physicists often talk about ``reduction'' in a way that suggests some theories, such as general relativity, are \emph{not} reduced, even though they will come out as reduced on our analysis.  At issue, we claim, is an ambiguity in the use of the word ``reduction''.  As we will argue, there are other uses of the term in physics with different meanings, and a theory may well be reduced in one sense but not others.  As one example, to illustrate the more general point, we consider the construction known as ``sympletic reduction'', often used by physicists as a precursor to quantization.  We will argue that symplectic reduction differs, both in motivation and outcome, from both what Dewar considers in his first-order logic examples and the considerations that seem to be at issue in the field theory cases.  We suggest that one source of confusion here is that similar tools in differential geometry are deployed in different ways in different parts of physics, leading to subtly different interpretations of symmetries in different applications.  Drawing on \citet{Bradley+Weatherall}, we will argue that in some applications of geometry to physics, ``representational redundancy'' leads to interpretive ambiguity, and so there are pragmatic reasons to remove that redundancy, even though it does not necessarily signal surplus structure nor is it the sort of thing that could be removed via sophistication.

The remainder of the paper will be structured as follows.  In section \ref{sec:sophistry}, we will present Dewar's account of symmetries, sophistication, and reduction in the first-order case.  We will highlight how these definitions have the key formal properties mentioned above.  Then, in section \ref{sec: analogies}, we will discuss problems that arise with extending these definition to the context of local field theories.  Section \ref{sec:electromagnetism} will work through the concrete example of electromagnetism in light of the foregoing, to show how even in established applications the distinction between reduction and internal sophistication is not as clear as one would like.  In section \ref{sec: sympreduction}, we will argue that reduction as characterized by Dewar's account and symplectic reduction are conceptually distinct. Finally, in section \ref{sec:auxiliary} we will argue that context-sensitive pragmatic considerations, rather than general arguments about symmetry and reduction in physics, should guide our choices between different theory (re)formulations, even in cases where a theory has been reformulated to remove redundancies and excess structure.  Section \ref{sec:conclusion} concludes.  

\section{Dewar on Symmetry, Reduction, and Sophistication}\label{sec:sophistry}

We begin by discussing Dewar's treatment of symmetry, sophistication, and reduction in the context of first-order logic.  Dewar motivates his account by considering a first-order theory with two one-place predicates, $L$ and $R$, and two axioms:
\begin{subequations}
\begin{align}
    \forall x&(Rx \vee Lx)\\
    \forall x\neg &(Rx \wedge Lx)
\end{align}
\end{subequations}
Models of this theory are structures consisting of a domain of objects and two subsets of that domain, with vanishing intersection, whose union is the entire domain.  As Dewar observes, we can think of this as a theory about worlds of gloves in which every glove is left-handed or right-handed, but none are both.

This theory has a symmetry, in the following sense: the theory is (non-trivially) translationally equivalent to itself.  What this means, in the present case, is that there are ``dictionary maps'' $\mathfrak{D},\mathfrak{D}'$ taking predicates of the theory to formulas of the theory such that for any formula $\varphi$, $T\vdash \varphi$ only if $T\vdash \mathfrak{D}\varphi$ and $T\vdash\mathfrak{D}'\varphi$, and for any formula $\varphi(x_1,\ldots, x_n)$, $T\vdash\forall x_1\cdots x_n (\varphi(x_1,\ldots x_n)\leftrightarrow \mathfrak{D}\circ\mathfrak{D}' \varphi(x_1,\ldots x_n))$ and likewise with $\mathfrak{D}$ and $\mathfrak{D}'$ reversed.  In Dewar's example, the symmetry is that we can translate the predicate ``R'' to ``L'', and vice versa, while satisfying the conditions stated.  This suggests that one can freely call gloves of one orientation ``right'', as long as one also calls gloves of the other orientation ``left'', and vice-versa.  It is a matter of convention which orientation gets which label.  Alternatively, one could argue that there is a kind of redundancy in the theory, in that there is no fact of the matter about which gloves are truly ``right'' and which are truly ``left''.  There are just two exhaustive and mutually exclusive categories of gloves.

Dewar goes on to note that some philosophers find symmetries of this sort dissatisfying.  (We will not comment on their reasons for dissatisfaction at present, but simply record the sociological fact.)  To resolve the dissatisfaction, he proposes two possible reactions.  One, he suggests, is already common in the literature: move to what he calls a ``reduced'' theory, which is a theory that plausibly can express the same facts about the world, but which does so without the use of arbitrary conventions.  To illustrate what this strategy amounts to, Dewar offers a candidate reduced theory: the \textit{congruence theory}.  This theory has a different signature, consisting of just a two-place relation, $C$, and the following axioms:
\begin{align*}
&\forall x(Cxx)\\
&\forall x\forall y(Cxy \rightarrow Cyx)\\
&\forall x \forall y\forall z((Cxy\wedge Cyx)\rightarrow Cxz)\\
&\forall x \forall y \forall z((\neg Cxy \wedge \neg Cyx) \rightarrow Cxz)
\end{align*}
In other words, $C$ is an equivalence relation that partitions the universe of gloves into at most two possible categories.  But unlike the handedness theory, it does so only relationally.

How should we think about the congruence theory?  There are a few remarks to make.  The first is that the congruence theory is not equivalent to the handedness theory, at least under the intended ``translation'' that takes, for any model of the handedness theory, the respective collections of left and right-handed objects to distinct cells of a congruence partition.   We can see this by considering a very weak notion of theoretical equivalence for first-order theories, which is \emph{categorical equivalence}.  

We define categories $\mathcal{C}$ and $\mathcal{H}$, for the congruence and handedness theories respectively.  For each of these, the objects are models of the theories and the arrows are isomorphisms.\footnote{ One could consider elementary embeddings, but since every isomorphism is an elementary embedding, the claim we make here is actually stronger.} The intended translation determines a functor $F$ that takes objects of $\mathcal{H}$ of the form $(D,L,R)$, where $D$ is some set and $L,R$ are disjoint subjects of $D$ containing the left and right-handed elements respectively, to objects of the form $(D,C)$, where $D$ is the same domain and $C$ is the set $\{<x,y>:(x,y\in L)\vee (x,y\in R)\}$, i.e., all ordered pairs of objects where both objects are left-handed or both are right-handed.  Arrows of $\mathcal{H}$ get mapped to arrows whose action on $D$ is the same.  This functor is not an equivalence of categories, because it fails to be full.  Consider, for instance, the model $(\{x,y,z\},\{<x,x>,<y,y>,<z,z>,<y,z>,<z,y>\})$ of $\mathcal{C}$.  This is a model with three objects, one of which is in one congruence class and the others of which are in the other.  This model is the image under $F$ of two models of $H$: one where there are two left-handed objects (and one right-handed object) and one where there are two right-handed objects (and one left-handed object).  Since those are not isomorphic as models of the handedness theory, there is no arrow of $\mathcal{H}$ that $F$ can map to the identity arrow on $(D,C)$.  Indeed, it turns out this functor is forgetful.  It \emph{forgets structure}, because its induced maps on arrows are not surjective \citep{Baez+etal,WeatherallUG,BarrettPSS}.  This inequivalence captures a sense in which the handedness theory has extra structure.

The fact that these theories are not equivalent means that the sense in which they ``say the same thing'' is attenuated.  The handedness theory can make assertions that the congruence theory cannot---namely, things about whether gloves are truly left-handed or truly right-handed.  This could matter for some purposes, such as in cases where the theory is embedded in some background linguistic practice where ``left'' and ``right'' have established meanings.  One might also be moved by Kantian considerations, and think that in a world where there is only one glove, there should still be some fact of the matter about its handedness.  The handed theory allows us to make sense of that possibility. It follows that worries of ``redundancy'' are not forced on us by the existence of the symmetry.  Rather, it is that there are situations in which we think the full expressive resources of the handedness theory are unnecessary.  In other words, it is the presence of the symmetry, plus a background understanding about what kinds of facts we think the theory should be able to express, that together support the move to a theory with less structure.

This point can be amplified by noting that the congruence theory also has a symmetry: namely, the one determined by the dictionary map that takes $C(x,y)$ to $C(y,x)$.  Dewar does not comment on this symmetry, but its presence suggests two possibilities.  On the one hand, one could say that the congruence theory is only \emph{partially} reduced; or else one could argue that the congruence theory is indeed a reduced theory, as claimed, but that reduced theories may still have symmetries.  The first option strikes us as non-viable. Any equivalence relation in first-order logic will have a symmetry of the sort described.  It seems to us that the congruence theory says precisely what Dewar wants it to say. It is a structural feature of congruence classes that they have precisely this symmetry.

This leaves the second possibility, which we take to be strongly consonant with the remarks above, namely, that whether a symmetry signals anything problematic depends on background commitments about what we intend the theory to express.  In the present case, we want to express that pairs of objects are congruent, and that the order in which the objects are presented is irrelevant to their congruence class. So we introduce axioms that enforce that within first-order logic.  The symmetry is a consequence of how we have built the formalism and the theory.  Further ``reduction'' would presumably remove the ability to express what we intend the theory to express.\footnote{ We note, for future reference, that there is another option, which is to say that although the transformation just described is a symmetry on Dewar's definition, it is a \textit{syntactic} symmetry, but not a \textit{semantic} symmetry, since its action on models is trivial.}

So much for reduction.  What about sophistication?  Here, Dewar offers the following: ``the idea is that we need not insist on finding a theory whose models are invariant under the application of the symmetry transformation, but can rest content with a theory whose models are isomorphic under that transformation'' (p. 498).  Dewar then gives several suggestions about what this might mean.  One is a perfectly clear recipe for constructing a sophisticated theory: one takes the theory with which one began, one identifies models that one wants to say are equivalent, or symmetry-related, and one introduces isomorphisms, by fiat, that correspond to those equivalences.  There are conditions that need to be satisfied to ensure this can be done consistently -- for instance, isomorphisms must be invertible -- but if those conditions are met, one can always generate a new ``theory'' in which the symmetries are implemented as isomorphisms.\footnote{ The requirements that need to be met are ones that ensure that the new isomorphisms, together with the old, form a groupoid.  This procedure was first described in these terms by \citet{WeatherallNGE}, though \citet{Halvorson+etal} suggest a similar strategy is effectively adopted by \citet{BelotNW}.  \citet{ChenSymmetry} provides a systematic treatment of this strategy with remarks on how to interpret it.}  This approach to sophistication has come to be known as ``external sophistication''.

From our perspective, external sophistication is well-defined and generally well-motivated.  It is analogous to a standard practice in mathematics, where one introduces a new kind of object through some quasi-axiomatic definition, often in the language of set theory, and then immediately defines a class of morphisms, including isomorphisms, that preserve the intended structure.  But some authors have objected to external sophistication.  For instance, \citet{Read+Martens} suggest that external sophistication is an instance of theft over honest toil, because one stipulates what maps preserve structure without giving an independent characterization of what structure is preserved.  External sophistication, they argue, obscures the ``ontology'' of the models.  To achieve our interpretive and perhaps even representational goals, one needs to understand sophistication in a different way.  One needs to say something about the ``internal'' structure of models.  As Read and Martens acknowledge, Dewar anticipates a worry of this sort, and though he discusses external sophistication as a strategy, the examples he actually gives of sophistication have a different character.  They are examples of ``traditional'' (Read and Martens) or ``internal'' (Dewar) sophistication.  

We will presently argue that there is no clear, general account of what internal sophistication amounts to.\footnote{As we noted in footnote \ref{fn:otherPeople}, other attempts to make ``internal sophistication'' precise have appeared since Dewar's paper.  But as we discuss in footnote \ref{fn:relativity} below, they are either subject to the same criticisms we make here, or else they change the subject.}  But to see the rough idea, consider how Dewar treats sophistication for the handedness theory.  There, he proposes to modify the semantics of first-order logic in order to make models isomorphic if they differ only with respect to which congruence class is called ``left'' and which ``right''.  He introduces a new kind of structure, apparently bespoke to the handedness theory, called a ``de-handed picture''.  A de-handed picture $m$ consists of a set $|m|$, representing the domain of the picture; a two-element set $\chi^m$; and a function $\epsilon^m:\chi^m\rightarrow \mathcal{P}(|m|)$, satisfying $\bigcup\epsilon^m[\chi^m]=|m|$ and $\bigcap\epsilon^m[\chi^m]=\emptyset$.\footnote{ Dewar does not mention these requirements, but they are needed to satisfy the axioms.}  De-handed pictures come with a new notion of homomorphism $h:m\rightarrow n$, consisting of a pair of maps $h=(h_1:|m|\rightarrow |n|,h_2:\chi^m\rightarrow\chi^n)$, such that for any $\star\in\chi^m$, $h_1[\epsilon^m(\star)]=\epsilon^n(h_2(\star))$.  An isomorphism is an invertible homomorphism (whose inverse is also a homomorphism).  This implements the idea that isomorphisms of de-handed pictures must preserve congruence classes, but need not preserve whether those classes are the extension of $L$ or of $R$.  Dewar goes on to describe a sense in which de-handed pictures can provide a semantics for precisely the ``symmetry-invariant'' assertions of the handedness theory. The end result is a class of mathematical objects -- in some extended sense, ``models'' of the handedness theory -- along with a class of homomorphisms and isomorphisms relating them.\footnote{ Dewar emphasizes that this process results in a theory categorically equivalent to the externally reduced one, though subsequent authors do not appear to take that to be a necessary feature of internal sophistication and so we do not dwell on it.}  One does this without introducing a new axiomatization for the theory, but rather by changing something about how the theory is ``interpreted''.  

So much for the motivating examples from first-order logic. We will now turn to classical field theories. As we will see, the analogies are not perfect, and ``sophistication'' will prove to be an unfortunate choice of terminology.

\section{Analogies and Permutations} \label{sec: analogies}

Dewar's ultimate goal in introducing the distinction between reduction and sophistication is to apply it to physical theories.  He focuses on local field theories---electromagnetism, Yang-Mills theory, general relativity, and so on.  But here he encounters a well-known challenge, which is that these theories do not have standard first-order axiomatizations.  Indeed, we do not usually think of classical field theories as having an associated ``signature'' of non-logical symbols (predicates), nor do we have formulas or even a ``theory'' in the senses needed to directly apply Dewar's definition of translational equivalence.  So instead of directly applying the ideas just discussed, Dewar works by analogy---and, to an extent, homophony.  But as we will presently argue, doing so leads to problems.

\subsection{Sophistication in the wild}

We first address an argument from homophony, mostly to set it aside.  We do not mean to say that Dewar places great weight on this argument, but it does seem to play a role in guiding his judgments---as he acknowledges \citep[p. 501]{DewarSoph}. As we noted in the Introduction, the most common usage of the term ``sophisticated'' in the foundations of classical field theories prior to Dewar's paper was in connection with a family of views collectively known as ``sophisticated substantivalism'' \citep{Belot+Earman,PooleyOld,PooleyOxford}.  These are views about the metaphysics of spacetime in general relativity characterized by the idea that spacetime substantivalism is compatible with ``Leibniz equivalence'', i.e., the doctrine that isometric spacetimes represent a single possibility.  (It is not entirely clear why Leibniz equivalence is meant to be incompatible with substantivalism, though \citet{Earman+Norton} influentially argued that the denial of Leibniz equivalence is a kind of acid test for substantivalist views.) One might initially imagine that sophistication in Dewar's sense is meant to be a generalization of whatever makes sophisticated substantivalism ``sophisticated''. 

But on reflection, it is difficult to see how that argument is supposed to go.  It is true that in general relativity, models of the theory are generically isomorphic to many other models, much like one finds in sophisticated theories in the first-order case.  These isomorphisms are sometimes called ``symmetries'', and sophisticated substantivalists deny that they are problematic.  But then again, the congruence theory also has many ``symmetries'' in this sense, as does any other candidate ``reduced'' first-order theory, since given any model of that theory, one can construct arbitrarily many different isomorphic copies, simply by permuting or substituting the elements of the domain and pushing forward the congruence relation along the permutation/substitution map.  Indeed, this is precisely how one generates isomorphic copies of models in general relativity. From this perspective, general relativity looks more like a reduced theory than a sophisticated one, per Dewar's usage. Or at least, it is strongly analogous to the congruence theory, in the sense that the transformations under consideration as ``symmetries'' are isomorphisms generated by permutations of a model domain, such as one finds in the congruence theory (or any theory), rather than ``additional'' maps that are not generated in this way, as in the sophisticated handedness theory.  

Is there perhaps a different sense in which general relativity should be seen as sophisticated?  We do not think so.  If we reflect on the examples of sophisticated theories that Dewar describes in the first-order case, the hallmark features are that one begins with some clearly articulated theory, one argues that it has a symmetry, and then one performs some surgery on the models of the theory -- either by adding new isomorphisms by hand, or by manipulating the models in such a way that new isomorphisms seem natural -- so that symmetry-related models turn out isomorphic. But that is not what happens with general relativity.  To the contrary, on the ``standard'' version of the theory, the one found in all modern textbooks, the models at issue are already isomorphic.  There is nothing to operate on. We neither introduce ``extra'' arrows nor do we manipulate the semantics of the theory in order to make non-isomorphic models isomorphic.\footnote{ That being said, one can certainly cook up alternative theories, such as the ``enriched'' general relativity discussed by \citet{WeatherallHoleArg} and \citet{Bradley+Weatherall}, where one takes ordinary general relativity and adds (uncountably many) names for all of the points.  This theory, if it were adequately regimented, presumably would have symmetries in Dewar's sense, since one would expect that provable claims about GR would be invariant under ``renaming'' points.  One could also imagine sophisticating this theory, or reducing it. But this theory is quite different from ordinary general relativity.}  

We conclude that sophisticated substantivalism does not have any salient relationship with sophistication in Dewar's sense---and that general relativity should not be seen as a sophisticated theory.  We also note that permutations should not be viewed as symmetries, at least not in the sense Dewar discusses for first-order theories; and reiterate the observation from above that isomorphisms between models do not signal the presence of symmetries or indicate any kind of ``excess structure''.

\subsection{Sophistication or reduction?}

So much for what sophistication is not.  How \emph{should} we understand reduction and sophistication in the context of local field theories?  We begin with reduction.  As in the first-order case, Dewar's idea is that if one has a theory with symmetries that one takes to signal some sort of redundancy, or excess structure, a reduced theory is a new theory that lacks those symmetries but otherwise expresses what we aim to express.  But we immediately encounter the problem noted above, which is that since we are not working with first-order theories, we do not have a general account of translational equivalence available. Instead, following Dewar, we apparently have to identify symmetries with transformations on the space (or category) of models of our theory.  For reasons we saw in the first-order case, this will be a coarser-grained notion of symmetry, since there are translational equivalences that will act as the identity on the space of models, just as $Cxy\rightarrow Cyx$ does.  But let us carry on anyway.\footnote{ Actually, Dewar defines symmetries for a special class of local field theories over $\mathbb{R}^n$, where the fields are all valued in a (fixed) copy of $\mathbb{R}^q$.  In this context, symmetries are diffeomorphisms from $R^n\times R^q\rightarrow R^q$ that induce bijections on the models (``solutions'') of the theory.  This is an effort to capture the ``syntactic'' part of translational equivalence, but the context in which it is defined is so limited that his own examples apparently cannot be fit into it, and so we simply set the definition aside and focus on its upshot.}

Since we do not attempt to axiomatize our theories, any candidate for a reduced theory will need to be presented as a category of models.  This means that we will work on what would have been the semantic side in the first-order case, by constructing new mathematical structures and defining salient notions of morphism between them.  Our goal is to come up with a theory that has less structure than the one we began with---and, in particular, which does not have problematic symmetry-related models.   More generally, what we want is a theory that expresses what we feel we need to express in order for the theory to be empirically adequate, without drawing any further distinctions that do not have any empirical significance or which we have reason to think are conceptually otiose.

But if that is reduction, what is sophistication?  External sophistication seems clear enough.  When confronted with a classical field theory that exhibits symmetries, in the closest sense possible to that in which the handedness theory admits symmetries, which is to say, it has distinct, non-isomorphic models that one thinks are equally well-suited to represent the same situations, then if the symmetries meet certain simple desiderata, one can simply build a category of models with new arrows corresponding to the symmetry transformation.  This is a way of ``removing structure'' that seems exactly the same as external sophistication in the first-order case.

But internal sophistication is harder to see.  Recall that in the first-order case, to find an internally sophisticated theory, one manipulates the semantics of the original theory to produce a category of models without the problematic symmetries.  That is, one describes, using the methods of ordinary mathematical physics, classes of models and morphisms between them, aiming to come up with candidates to make symmetry-related models isomorphic.  But that is precisely what we just said reduction was for local field theories.  Is there a difference?  Certainly, unlike in the first-order case that Dewar considers, the difference \emph{cannot} be a difference between identifying new axioms and defining a new semantics for your old axioms, because there are no axioms to change (or preserve).  And without that distinction, we do not see any reason to distinguish reduction from internal sophistication.\footnote{\label{fn:relativity} Having made these arguments, we now substantiate the claim made in footnote \ref{fn:otherPeople}, regarding more recent attempts to articulate the difference between reduction and internal sophistication.  We focus on \citet{March+Read}, because their stated aim (which we take to be successful) is to offer mathematically precise definitions that track the distinction in the recent literature.  On their view, sophistication and reduction are both ``four-place relations between two theories, a map between those theories, and a set of symmetries of the former theory'' (1). The difference concerns whether the map between the theories takes the symmetries to identity maps on models, or to isomorphisms between models.  The core point, from our perspective, is that both are defined as \emph{relations} between theories.  But no such definition can perform the role that Dewar sets out for reduction or sophistication.  The reason is that whether a theory has a symmetry is apparently an intrinsic property of the theory (or, perhaps, a relation between the theory and its representational targets in the world).  Whether it is problematic that the theory has a symmetry is presumably also intrinsic.  So the problem Dewar seeks to solve is an intrinsic one.  We (charitably, we think) interpret him as offering an intrinsic solution.  March and Read, along with other recent authors, re-interpret sophistication and reduction as relational between theories.  But if that is what sophistication and reduction are, they have no chance at solving the problem for which they were originally deployed. (In section \ref{sec:auxiliary}, we will give a different reason to be skeptical of the sort of proposal March and Read advance, which is that it is pragmatic considerations, not a formal property of the symmetries of a theory [instrinsic, or relational] that should guide how we choose among formulations of theories.)}

\section{Dewar's Example} \label{sec:electromagnetism}

We conclude from the foregoing that there is no interesting difference between internal sophistication and reduction for classical field theories.  This might give some readers pause.  After all, in the years since Dewar's paper appeared, several other authors have drawn similar distinctions between internally sophisticated and reduced classical field theories.  Some have even given examples of distinct theories that are supposed to differ in that one is a reduced theory and the other is an internally sophisticated one, both relative to some initial theory that has symmetries one wishes to remove.\footnote{ In addition to the standard example discussed above, \citet{Lu+etal} analyze a half dozen formulations of teleparallel gravity and classify them on whether they are sophisticated or reduced.} How could this be, if there is no interesting difference between internally sophisticated and reduced theories?  

Dewar gives an example, based on earlier work by \citet{WeatherallNGE,WeatherallUG} on formulations of electromagnetism, that many authors have taken to be emblematic of the difference. For simplicity (and following Dewar), we will work on Minkowski spacetime.  Consider first a formulation of Maxwell electromagnetism based on vector potentials.  The models of this theory are triples $(M,g_{ab},A_a)$.  (For convenience, we will consider the vacuum sector of the theory, so we assume $g^{mn}(\nabla_m\nabla_n A_a -\nabla_m\nabla_a A_n)=\mathbf{0}$.  Nothing turns on this; we could also introduce sources, or define sources via satisfaction of Maxwell's equations.)  Following previous treatments, the natural choice of isomorphisms here would be diffeomorphisms $\chi:M\rightarrow M$ that preserve both $g_{ab}$ and $A_a$.  The empirical significance of the theory, meanwhile, appears to run through the effects that the vector potential has on matter.  At least classically, these can be described, for small point particles with mass $m$, charge $q$, and 4-velocity $\xi^a$, via the Lorentz force law: $m\xi^n\nabla_n\xi^a = q F^a{}_{n}\xi^n$, where $F_{ab}=d_a A_b$ is the Faraday tensor.\footnote{ Here $d$ indicates the exterior derivative.}  For more realistic, extended charged matter the relationship between the vector potential and accelerated motion is more complicated, but it, too, involves only the derivative of the vector potential, i.e., the Faraday tensor, and not the vector potential itself.

The theory just described is often thought to have a symmetry.  The reason is familiar: different vector potentials can give rise to the same Faraday tensor.  This is because there exist one-forms $G_a$ such that $d_aG_b=\mathbf{0}$.  (Such fields are said to be \emph{closed}; in particular, every field of the form $G_a=d_a\varphi$ will have this property, for any smooth scalar field $\varphi$.)  Thus, given any vector potentials $A_a$ and any closed one-form $G_a$, we can define $A'_a = A_a + G_a$, and then we find that $F_{ab}=d_a A_b = d_a A'_b = F_{ab}'$.  Since the empirical significance of the vector potential always runs through its associated Faraday tensor, it follows (again, classically) that no possible experiment could distinguish models $(M,g_{ab},A_a)$ and $(M,g_{ab},A'_a)$ differing in this way.  For Dewar, this should be thought of as a kind of redundancy, analogous to the redundancy in the handedness theory: we have multiple models that, as far as our intended applications go, are equally good for all purposes, and yet, by the lights of the theory, are not only distinct but inequivalent.  Thus the vector potential theory is a candidate for reduction or sophistication. 

As we have discussed, there is a perfectly well-defined procedure to externally sophisticate this theory: we stipulate, as additional theoretical content, that models related by a ``gauge transformation'', i.e., any transformation of the form $A_a\mapsto A'_a = A_a + G_a$ for $G_a$ closed, are ``isomorphic''.  That is, we move from the category of models with which we began to one with a new definition of arrows that includes, in addition to the diffeomorphisms already mentioned, these additional arrows.\footnote{\citet{WeatherallNGE} shows that these ``extra'' maps satisfy the conditions needed to form a new category, so we will not pause over that issue here.}  This gives a well-defined category of models, and it also gives a description of the intended physical content of each model as that structure preserved by the specified maps.  

Dewar offers two other reactions to this symmetry, which he identifies as reduction and internal sophistication, respectively.  The reduced theory modifies the models as follows: we now define models to consist of triples $(M,g_{ab},F_{ab})$, and we impose a further ``law'' stating that $d_aF_{bc}=\mathbf{0}$.  (Since we are working in Minkowski spacetime, and the underlying manifold is contractible, this law implies that there exists a covector field $A_a$ such that $F_{ab}=d_aA_b$.  The restriction to the vacuum sector now means $\nabla_aF^a{}_b=\mathbf{0}$.) The arrows associated with this theory are diffeomorphisms $\chi:M\rightarrow M$ that preserve $g_{ab}$ and $F_{ab}$.  In essence, we take each model of the original theory and we replace the vector potential with its associated Faraday tensor.  

This new theory removes the symmetry we identified before, since the models related by (just) a gauge transformation agree on their Faraday tensors, and so the equivalence class of distinct but empirically indistinguishable models all map to the same Faraday tensor.  Still, there is a sense in which this theory still has \emph{some} symmetries. (For instance, if we allow the sort of ``syntactic'' symmetries Dewar considers in the first-order case, we might identify $F_{ab}\mapsto -F_{ba}$ as a symmetry of the theory---though of course, that transformation acts trivially on models.)  Even so, this theory has ``less structure'' than the vector potential theory, according to the criterion given by \citet{WeatherallUG}, because distinct, non-isomorphic vector potentials map to the same Faraday tensors, and in this sense it removes symmetries from the theory with which we began.

The internally sophisticated theory Dewar proposes uses different machinery: it recasts electromagnetism as a special case of (Abelian) Yang-Mills theory.\footnote{ See \citet{Palais}, \citet{Bleecker}, \citet{WeatherallFBYMGR}, and \citet{GomesElement}.}  We still work in Minkowski spacetime, but now we introduce, over Minkowski spacetime, a (trivial) principal bundle $P\xrightarrow{\pi} M$, with structure group $U(1)$, endowed with a principal connection $\omega_{\alpha}$.  (Though $\omega_{\alpha}$ is technically a Lie-algebra-valued 1-form, the Lie algebra associated with $U(1)$ is $\mathbb{R}$, and so we drop the Lie-algebra-valued index on the principal connection.)   This connection is required to satisfy the Yang-Mills equation, which in the vacuum sector becomes $\star d_{\alpha}\star \Omega_{\beta\kappa}=\mathbf{0}$, where $\star$ is the Hodge star operator for horizontal $k-$forms on $P$ determined by the Minkowski metric on $M$, $d$ is now the exterior derivative on $P$, and $\Omega_{\alpha\beta}$ is the curvature 2-form associated with $\omega_{\alpha}$. We can write models of this theory as pairs $(P, \omega_{\alpha})$.  Symmetries of these models will be principal bundle automorphisms that preserve both the Minkowski metric and the connection, which is to say, pairs $(\Psi,\psi)$, where $\Psi:P\rightarrow P$ and $\psi:M\rightarrow M$ are diffeomorphisms satisfying $\pi\circ \Psi=\psi\circ\pi$, $\Psi(x\cdot g)=\Psi(x)\cdot g$ for any $g\in U(1)$, $\psi^*(\eta_{ab})=\eta_{ab}$, and $\Psi^*(\omega_{\alpha})=\omega_{\alpha}$.

This theory is related to the vector potential theory as follows.  Given any (global) section $\varphi:M\rightarrow P$, the pullback of the connection along $\varphi$ is a 1-form on $M$, $\varphi^*(\omega_{\alpha})=A_a$.  The exterior derivative is the pullback of the curvature of that connection, $F_{ab}=d_aA_b = \varphi^*(\Omega_{\alpha\beta})$.  And finally, satisfaction of the Yang-Mills equation implies that $F_{ab}$ satisfies the vacuum Maxwell equation.  So, given a model of the internally sophisticated theory $(P,\omega_{\alpha})$ and a section $\varphi:M\rightarrow P$, one can generate a model of the original, unsophisticated theory.  

In what sense has this new theory ``removed'' symmetries?  There are three key observations.  The first is that given a connection $\omega_{\alpha}$, and two sections $\varphi,\varphi'$, it is always the case that $d_a(\varphi^*(\omega_{\beta})-\varphi'^*(\omega_{\beta}))=\mathbf{0}$---that is, the pullback of $\omega_{\alpha}$ along different sections is always related by a gauge transformation.  Likewise, given any two gauge-related vector potentials, there always exists a principal connection and a pair of sections such that those two vector potentials arise as the pullbacks of those sections.  The final observation is that if we have two different connections $\omega_{\alpha}$ and $\omega'_{\alpha}$ such that, for a single section $\varphi$, we have $d_a\varphi^*(\omega_{\alpha}-\omega_{\beta})=\mathbf{0}$, then there is a principal bundle automorphism $(\Psi,1_M)$ such that $\Psi^*(\omega_{\alpha})=\omega'_{\alpha}$.   Thus, we can see different gauge-related vector potentials as representations, relative to different choices of section, of an invariant object, viz., the principal connection $\omega_{\alpha}$; or, equivalently, as symmetry-related principal connections expressed via a single section.\footnote{ There is another way of thinking about this relationship, which may be more enlightening for some readers, but which requires technical tools that we do not introduce here.  Fix a section $\varphi:M\rightarrow P$.  Then given any vector potential $A_a$ on $M$, we can determine a principal connection on $P$ via a two step process, where first we push $A_a$ forward along $\varphi$ to $\varphi[M]$, the image of the section as a submanifold of $P$, and then we impose what is known as the ``equivariance'' condition to extend this field on $\varphi[M]$ to all of $P$.}

This new theory has less structure than either of the formulations of electromagnetism discussed thus far.  This is because, in addition to the transformations on $P$ that correspond to transformations between gauge-transformation-related vector potentials, the structure $(P,\omega_{\alpha})$ is invariant under \emph{global} $U(1)$ transformations that leave the pullback of $\omega_{\alpha}$ along any section invariant.\footnote{ See \citep{Bradley+Weatherall} for discussion.  Our position here is somewhat different, insofar as we emphasize the problems with constant gauge transformations.} Hence, the natural functor from vector potentials (or Faraday tensors) to principal connections on $P$ is not full. It is true that an analogue of these global gauge transformations can be introduced to the vector potential formalism, by defining gauge transformations as smooth scalar fields $\varphi$ which act on vector potentials as $A_a\mapsto A_a + d_a\varphi$.  In that case, constant scalar fields yield constant gauge transformations.  But doing this is at best awkward and arguably incoherent, because it involves stipulating a class of transformations whose action on all objects is trivial, and yet which are counted as distinct transformations, simply for the purpose of restoring a kind of structural equivalence.  It is hard to see what is gained by this procedure, except to give an intuition about what the ``extra'' transformations in the principal bundle formalism accomplish.  

In any case, even if one does this, it does not truly resolve the issue.  The reason is that the new transformations can be seen as isomorphic to $\mathbb{R}$, as a Lie group; whereas the constant gauge transformations in the principal bundle formalism are isomorphic to $U(1)$, which is not isomorphic, as Lie groups, to $\mathbb{R}$, even though they have the same Lie algebra.  The result is that, while one can (artificially) redefine the theory of vector potentials so that there exists a functor from vector potentials to principal connections on $P$ that is full, thereby removing the concern that the principal bundle formalism should be seen as having less structure than vector potentials, the cost of doing so is that same functor fails to be faithful (because it winds $\mathbb{R}$ around $U(1)$ infinitely many times).  In any case, the theories are not equivalent.  We suggest that this modification of the vector potential formalism should simply be dismissed.

We now return to the question with which he began the section. Why think of the principal connection theory as ``sophisticated'' and the Faraday tensor theory as ``reduced''?  We see no compelling reason to distinguish them.  One might say that if the Faraday tensor theory is reduced, then if anything the principal connection theory is \emph{more} reduced, in the sense that it removes further structure.  On the other hand, since we have not given any axiomatizations, \emph{all} of our definitions and modifications of theories have involved manipulations on the semantic side, i.e., by building new mathematical structure by hand using the tools of ordinary mathematics.  From this perspective, both the Faraday tensor and principal connection theories are ``internally sophisticated''.  So we can see reasons to classify both theories as reduced, or both as sophisticated---but not to draw a distinction between them.

We acknowledge that there are some formal differences between the theories that seem to guide intuitions about how to classify these theories.  For instance, the formulation using the principal connection has distinct, non-trivial symmetries that, in a certain precise sense, ``correspond to'' the gauge transformations introduced in external sophistication, via the functor with which we have been comparing the theories.  (Of course, it also has additional symmetries, for reasons we have discussed.)  The Faraday tensor theory, meanwhile, does not.  While it does have non-trivial isomorphisms, corresponding to spacetime isometries that preserve the Faraday tensor, there is a sense in which all ``pure'' gauge transformations are mapped to the identity.  

This is a real difference, but we do not see any reason to put conceptual weight on it.\footnote{As we discuss in section \ref{sec:auxiliary}, there are other differences between the theories as well, but they have nothing to do with reduction and sophistication.}   As we discussed above, in connection with ``sophisticated substantivalism'', and as has been observed by many others, \emph{any} sufficiently rich theory of physics formulated using standard mathematical tools is going to have models with isomorphic copies; and generally, one should expect some models to have non-trivial automorphisms.  That is equally true of both the Faraday tensor formulation and the principal bundle formulation---just as it is true of general relativity and the vector bundle formulation of electromagnetism.  In all of these cases, one must be ``sophisticated'' in the sense of sophisticated substantivalism, and recognize that the intended structure is whatever model structure is preserved by model isomorphisms.  But as we have already emphasized, this sort of sophistication is simply orthogonal to the kind of sophistication apparently at issue in the first-order case, or as applied in external sophistication.  

\section{Reduction, Revisited} \label{sec: sympreduction}

We conclude from the foregoing that Dewar's own example does not establish an interesting difference between reduction and (internal) sophistication.  But perhaps the problem is the example.  After all, there are other examples in physics where something called `reduction' is contemplated.\footnote{As noted in the introduction, we offer symplectic reduction as just one example of a form of ``reduction'' that is conceptually distinct from what Dewar considers.  Our point is that the fact that physicists often talk about reduction does not imply that there is a sharp or meaningful distinct between what Dewar calls ``reduction'' and ``internal sophistication'', because of the polysemy of ``reduction''.}  Consider, for instance, symplectic reduction: one formulates a theory on a phase space, and then one quotients out the (gauge) symmetries of the theory by defining a new state space whose points consist of the equivalence class of points along the orbits of the symmetry group. Perhaps in that context we can draw a sharper distinction between sophistication and reduction?

We think the answer to this question is ``yes''.  But some care is needed, because the \emph{reason} a sharper distinction is available is that symplectic reduction bears little resemblance to the sort of reduction that Dewar contemplated in the first-order case, where one introduces a new theory, with different syntax and axioms, that expresses only the ``invariant'' facts of a theory with translational auto-equivalences.  (In what follows, call reduction in Dewar's sense ``D-reduction'', to distinguish it from sympletic reduction or other procedures called ``reduction'' in the literature.)  

We acknowledge that there are some ways in which symplectic reduction resembles D-reduction. In both cases, one starts with a theory with symmetries that one takes to signal a sort of redundancy, and one constructs a theory that lacks those symmetries. Moreover, it seems to capture a stronger sense of ``removing" symmetries than simply making a collection of symmetry-related models isomorphic, since the distinction between the points related by the relevant symmetries on the original state space collapses through symplectic reduction. That is, it seems to capture precisely the idea of characterizing a theory \emph{only} in terms of the ``invariants" under some symmetry group.  

But there are crucial differences. For one, D-reduction is supposed to apply to any theory with symmetries (though saying what ``symmetry'' means in a sufficiently general way proves difficult).  Symplectic reduction, meanwhile, applies only to theories formulated in a certain way --  on a symplectic, or presymplectic manifold -- and with a specific class of symmetries. Symplectic reduction is therefore at best a special kind of reduction. 

But more importantly, D-reduction is a procedure for removing excess structure from a theory, i.e. modifying a theory with non-isomorphic models, each representing putatively distinct physical situations, but which one takes to be physically equivalent. Symplectic reduction works differently.  For one, physical situations, which we previously understood to correspond to models of a theory, are now represented as points of a manifold.  The formalism does not encode any additional information about equivalence between the points of the manifold, since the points have no ``internal structure'' that could be equivalent or inequivalent.  (Or, alternatively, every point is equivalent to every other point.)  What previously was represented ``internally'' to a model is now externalized to the symplectic structure on the manifold.  

Meanwhile, in this context the natural notion of ``symmetry'' is smooth maps from the (symplectic) manifold of configurations to itself that preserve the symplectic structure and a Hamiltonian function.  This notion of symmetry has no straightforward relationship to dictionary maps or autotranslations (there is nothing to translate); and one cannot ``remove" symmetries of this sort by manipulating the internal structure of configurations, \`a la internal sophistication and D-reduction as discussed above, because the points of the manifold have no structure to manipulate.  Instead, one acts globally on the manifold.

These remarks are best illustrated by an example. Consider symplectic reduction in the context of Hamiltonian formulations of gauge theories, where by ``gauge theory'' we now mean a theory  formulated on a manifold $M$ (the ``constraint surface" in phase space) whose points consist of the possible generalized position and canonical momenta of the system that satisfy some collection of constraints, $\varphi_i=0$. $M$ is equipped with a presymplectic two-form $\tilde{\omega}$ and a Hamiltonian function $H$, and the dynamics of the system is specified according to $\tilde{\omega}(X_H,\cdot)=dH$. We also define a collection of ``gauge orbits" on $M$ that consist of all the points connected by the integral curves of the null vector fields of $\tilde{\omega}$.\footnote{ For further details on gauge theories in the constrained Hamiltonian formalism, see \citep{henneaux}.}

The gauge orbits are often taken to indicate a kind of ``redundancy" in the theory: they correspond to distinct points of the constraint surface that represent the same physical situation. Given that we do not want to mistakenly treat these points as corresponding to distinct possible configurations, it is often thought that we should reformulate the theory so that instead of being formulated on $M$, it is formulated on some quotient manifold $\bar{M}$ (the ``reduced surface") whose points consist of the equivalence classes of points along the gauge orbits on $M$. A symplectic two-form is induced on the reduced surface, and the dynamics on the reduced surface are well-defined and give rise to unique solutions.\footnote{ For further discussion of symplectic reduction and when it is well-defined, see \citep{belotsymmetry}. }

However, a gauge theory (defined as above) is not a theory with excess structure in the sense of having non-isomorphic models that one takes to represent equivalent situations. The reason is that although there appears to be a kind of ``redundancy" in the theory, the transformations along the gauge orbits are (non-trivial) automorphisms of the models of the theory---understood as the structure $(M, \tilde{\omega}, H)$---and so are already isomorphisms in the theory.\footnote{ The reason is that the transformations along the gauge orbits on the constraint surface are diffeomorphisms that preserve the pre-symplectic two form $\tilde{\omega}$ and the Hamiltonian function on the constraint surface (see \cite{bradleygauge}).} 

The fact that the theory on the constraint surface is not a theory with excess structure suggests that the type of redundancy at issue is not the same as in the context of the sophistication vs. reduction literature. Instead, the type of redundancy at issue is one that \citet{Bradley+Weatherall} call ``representational redundancy": one can use a particular mathematical object in lots of different ways to represent the same physical situation via its non-trivial automorphisms. In the example above, such representational redundancy leads to an interpretive ambiguity regarding the points along the gauge orbits: they are represented as distinct points and yet they are structurally equivalent with respect to the symplectic structure. This might give rise to pragmatic reasons to remove that redundancy via symplectic reduction since symplectic reduction equivocates between the points along the gauge orbits. For example, one of the motivating reasons for formulating a gauge theory on a symplectic reduced space is to have a well-posed initial value problem. However, it does not signal surplus structure, nor is it the sort of thing that could be removed via sophistication. Moreover, as \citet{Bradley+Weatherall} argue, removing this kind of redundancy can often lead to \textit{adding} structure. 

One might try to respond that we should think of the constraint surface theory in the example above as the ``internally sophisticated" version of some \textit{other} theory with excess structure, such that the symplectic reduced theory is the ``D-reduced" version of this other theory. For example, we might take the theory with excess structure to be formulated on the unconstrained manifold, since arbitrary gauge transformations will not necessarily preserve the symplectic two-form on phase space. Indeed, one might say that it is exactly the difference between the constraint surface theory and the symplectic reduced theory that characterizes the difference between ``internal sophistication" and ``D-reduction". But why should we call the symplectic reduced theory the D-reduced theory and not the constraint surface theory? As we pointed out above, there is a sense in which the symplectic reduced theory has \textit{more} structure than the constraint surface theory, relative to some functor between the associated categories.

To further emphasize the difference between symplectic reduction and D-reduction as ways of capturing a theory without excess structure, it is helpful to contrast symplectic reduction with structure excision in the context of spacetime theories. Take, as an example, Newtonian spacetime, which we can consider as a category whose objects are given by models $(M, t_{ab}, h^{ab},\nabla, \xi^a)$ and arrows are given by the diffeomorphisms on $M$ whose pushforward preserves $t_{ab}, h^{ab},\nabla$ and $\xi^a$. Newtonian spacetime has excess structure in the sense that the laws of Newtonian gravitation do not depend on the structure $\xi^a$---or, to put it in terms familiar from previous sections, the spacetime has a ``symmetry'' in the sense that there are equivalences between models that are not isomorphisms. The standard way of removing this excess structure is to move to Galilean spacetime $(M, t_{ab}, h^{ab},\nabla)$, whose automorphisms are the diffeomorphisms on $M$ whose pushforward preserves $t_{ab}, h^{ab},\nabla$. (This process can be thought of as D-reduction, or sophistication.) 

Suppose, instead, that we applied the sort of reasoning that motivates symplectic reduction: we note that there are points of Newtonian spacetime that are ``structurally equivalent'' in the sense of being related by diffeomorphisms that preserve the relevant structure, and we identify the points related by these non-trivial automorphisms by passing to a manifold of equivalence classes of points under these symmetries.  We claim this strategy would be absurd---and would neither remove structure nor result in a coherent representation of spacetime.\footnote{ That does not mean this sort of idea is never considered in the literature---see the debate between \citet{Wuthrich} and \citet{Muller}.} The reason is that the points related by these automorphisms are intended to represent distinct but structurally equivalent locations at which events can occur in space and time.  It is generally important for our representational aims to be able to represent them as distinct. Passing to orbits under the symmetries of Newtonian spacetime would deprive us of the resources we use to do so, because points representing distinct locations would be identified.  Indeed, we can press the point even further: Galilean spacetime, where the excess structure of Newtonian spacetime has already been removed, has even more automorphisms, and thus even more points that would be identified under the contemplated scheme.

What is going on?  The issue is that the non-trivial automorphisms in the spacetime case do not give rise to representational ambiguity in the same way as in the case of the non-trivial automorphisms of phase space. The points of the constraint surface in phase space are supposed to represent entire configurations, and so the natural interpretation of the constraint surface is to take distinct points to represent distinct configurations. But the non-trivial automorphisms are such that they move around the points of the constraint manifold, and so we want to be able to think of some points as representing equivalent configurations. This motivates one to ``equivocate" between those points of phase space representing equivalent configurations, and thereby remove the associated representational redundancy. On the other hand, it is not problematic that one can represent the same physical situation using different points of a space-time manifold, since these points do not have the same representational significance as the points of phase space.\footnote{ The cases where this \textit{is} problematic are where one has some background commitment to relationalism, such that one does not want to formulate a spacetime theory on a spacetime manifold at all. For example, \citet{belotsymmetry} argues that a viable relationist strategy is to take the phase space formulation of a space-time theory and symplectically reduce the theory under the non-trivial automorphisms of spacetime.}


Taking stock, we have argued that symplectic reduction is (sometimes) a well-defined procedure that removes certain kinds of symmetries---non-trivial automorphisms---from a physical theory. However, the sorts of considerations that motivate symplectic reduction and that motivate excess structure excision are importantly distinct, as are the results of applying each method. Although similar tools from differential geometry are deployed in both cases, the interpretation of symmetries in each differs. And as we have argued, symplectic reduction does not in general give rise to a theory that only removes excess structure. It therefore cannot provide the analogue of the congruence theory in the handedness example, and consequently, it does not provide a way of capturing D-reduction.


\section{Pragmatic Considerations} \label{sec:auxiliary}

Where does this leave us? On the one hand, we have argued that there is no meaningful distinction between ``reduction" and ``internal sophistication" in the context of physical theories, and that ``symplectic reduction" means something different to either of these. On the other hand, there do seem to be important differences between different formulations of theories that have been concocted to remove structure---including differences between formulations that have been labelled ``sophisticated'' vs. ``reduced'' in the literature. Indeed, as we mentioned in the introduction, some authors, Dewar included, have argued that ``sophisticated" theories can have explanatory advantages over their reduced alternatives. Is there some alternative distinction that grounds the intuition that there is a meaningful difference between theories that the terms ``sophisticated'' and ``reduced'' were supposed to capture? 

Ultimately, we think that any such attempt will have to take into account pragmatic and context-dependent considerations, rather than a formal difference of the sort the reduction/sophistication distinction aims for.\footnote{Background philosophical commitments might also matter, such as when 17th century physicists preferred relationist or mechanist theories on apparently \emph{a priori} grounds. We are neutral on whether these are ``pragmatic'' considerations (e.g., a demand for intelligibility) or non-pragmatic, but we allow that they do sometimes matter in practice, and simply note that they, too, do not track formal differences of the sort at issue here.}  The terms ``sophistication" and ``reduction" at best track some of the heuristic considerations connected to symmetry and structure that go into formulating a physical theory, but these considerations should not be taken to be universal, nor can they be defined on purely formal grounds. 

One way to see some of the pragmatic reasons for removing structure in one way over another draws from a distinction made from \citet{BradleySoph}. Bradley argues that there are representational advantages to those theories that Dewar, and others, have characterized as sophisticated, which she captures through a notion that she calls ``auxiliary structure".  Roughly, the idea is that although the \textit{invariant} content of a sophisticated and reduced theory is the same (what \citet{BradleySoph} calls the ``theoretical structure"), there might be some quantities that one can distinguish in a model of the sophisticated theory that one cannot distinguish in the corresponding reduced model. The ability to distinguish such quantities is attributed to the fact that the sophisticated theory has some ``auxiliary structure" that the reduced theory lacks. 

Although \citet{BradleySoph} takes for granted the distinction between sophistication and reduction that we have criticized here, we think that the notion of ``auxiliary structure'' can still be valuable. The reason is that we can understand having more auxiliary structure not as tracking a general formal property of a theory -- as the notions of ``sophisticated'' and ``reduced'' are supposed to do -- but rather as tracking a (context-relative) difference that can be drawn between theories that may be pragmatically important.   

To see this in further detail, let us consider the example of Maxwell electromagnetism again. We argued in section \ref{sec:electromagnetism} that the differences between Faraday tensor and principal connection formulations of electromagnetism do not seem to correspond to the differences between internal sophistication and D-reduction in the first-order case. But as we also acknowledged, there is still a sense in which Yang-Mills theory has non-trivial maps that ``correspond to" the gauge transformations and that can be introduced through external sophistication, while the Faraday tensor formulation does not. Indeed, the Faraday tensor formulation, although categorically equivalent to the externally sophisticated theory, is not categorically isomorphic to it. 

One way to interpret this situation is that there are multiple distinct isomorphic models in the externally sophisticated theory and principal bundle formulation that correspond to identical models of the Faraday tensor formulation. The distinctness between these models is just what \citet{BradleySoph} aims to capture via the notion of auxiliary structure; one can distinguish the models by fixing some external standard of comparison, whereas no such option is available for the corresponding models of the Faraday tensor formulation. Calling the principal bundle theory ``sophisticated" might be seen to track this difference.   That being said, electromagnetism is also illustrative because, in fact, the Faraday tensor formulation has more structure than the principal bundle formulation. This turns out to be a barrier to applying it in certain cases with non-trivial topologies (see \citet{Nguyen+etal}). So we might alternatively view the principal bundle formulation as having representational benefits on the basis that it has less structure than the Faraday tensor formulation. Relative to this comparison, the difference between Faraday tensors and principal bundles is not that one is D-reduced and the other is not. It is that Faraday tensors have too much structure to generalize to other applications of interest. 


Similarly, take the example of a constrained Hamiltonian theory formulated on the constraint surface and its corresponding symplectic reduced theory. Notice that while the gauge variables are well-defined quantities on the final constraint surface, they are not on the reduced surface, since by definition, the reduced theory identifies points of the constraint surface where the values of the gauge variables differ. So we might say, in the terms of \citet{BradleySoph}, that the theory on the final constraint surface has important auxiliary structure that the reduced theory lacks. And again, there might be pragmatic reasons for formulating a theory where one can define gauge variables, even if the points where the gauge variables differ are equivalent (in the sense of being related by an isomorphism). For example, Carlo \cite{rovellipartial,rovelliwhy} has argued that gauge variables are important for representing coupling between subsystems. 


This highlights that the fact that a given symmetry group acts as non-trivial isomorphisms in one theory but as the identity in another might be a relevant difference between the two, given some representational concerns. The way the terms ``(internal) sophistication" and ``D-reduction" have been applied in the literature suggests that they are often employed to capture such pragmatically significant differences between theories. The issue is that this fails to align with the characterization of these terms---namely, as denoting conceptually distinct methods for removing excess structure from a physical theory. In the examples above, the theory that has more auxiliary structure also removes structure from the theory with less auxiliary structure, and so is \textit{more} D-reduced. And so again, we suggest dropping the terms ``internal sophistication" and ``(D-)reduction", because they do not track a clean distinction.  But that does not mean that there are no valuable distinctions to draw, only that to properly track those distinctions, we need to consider context-sensitive pragmatic factors about how we intend to use our theories rather than general considerations about symmetry.

\section{Conclusion}\label{sec:conclusion}

Given a theory of physics with distinct, non-isomorphic models, there seem to be two strategies available.  One is to ``manually'' add transformations---the external approach.  The other is to manipulate models (always on the semantic side) to construct new structures that somehow correspond to, or can be constructed from, the models with which one began, but such that equivalent models end up being isomorphic.  These are different strategies.  The first is unique and generally well-defined, but it is plausibly unsatisfying.  The second is more amorphous, because it requires some judgment and because more than one theory can result.   


This second strategy has some features in common with reduction in the first-order case, and some features in common with internal sophistication.  But we do not see any conceptual advantages to trying to distinguish between these as different implementations of an ``internal'' approach to theory reformulation. 
The distinction in the literature between reduced and internally sophisticated theories, though it does in some cases reflect formal properties of theories, does not track any conceptually important distinction between ways of removing excess structure. For this reason, we suggest retiring the distinction between reduction and internal sophistication. 

The broader lesson that we want to take away from this discussion is that whether a symmetry signals redundancy, or the possibility of ``reduction" of this symmetry, depends on what one's background needs are. In some cases, a theory may possess symmetries that signal excess structure, while simultaneously lacking other symmetries that indicate that it is already reduced. In such cases, the meaningful question is not whether the theory is \textit{really} reduced, but rather which among the available theories best serves the explanatory or representational purposes at stake. It is therefore unsurprising that one is unable to capture a well-defined distinction between ways of removing excess structure that depends only on formal properties of theories. This is because what theories one takes to be viable alternatives, and how one interprets the relationships between these theories, depends on context-relative considerations concerning the role that different parts of a physical theory play.

\section*{Acknowledgments} JOW: This material is based upon work supported by the National Science Foundation under Grant No. 2419967. I am grateful to my Spring 2023 graduate seminar for very helpful discussions connected to the ideas in the paper; to Thomas Barrett's Spring 2023 seminar on symmetry and structure for feedback on a presentation related to this paper. CB: I am grateful to the audiences in Oxford, Bristol, and Cambridge for feedback on talks related to the paper. Both authors are grateful to Dave Baker, Thomas Barrett, Neil Dewar, Henrique Gomes, Caspar Jacobs, Eleanor March, and James Read for correspondence and discussion about the paper, including comments on previous drafts. We also thank helpful reviewers for their contributions to the final version.

\bibliography{sophisticated}
\bibliographystyle{elsarticle-harv}

\end{document}